# Optical Properties of Manganese Doped Wide Band Gap ZnS and ZnO

M. Godlewski[1,2], A. Wójcik-Głodowska[1], E. Guziewicz[1], S. Yatsunenko[1], A. Zakrzewski[1],

Y. Dumont[3], E. Chikoidze[3], M.R. Phillips[4]

[1]Inst. of Physics, Polish Academy of Sciences, Al. Lotników 32/46, 02-668 Warsaw, Poland

[2]Dept. Mathematics and Natural Sciences College of Science, Cardinal S. Wyszyński University, Warsaw, Poland

[3]Groupe d'Etudes de la Matiere Condensee, CNRS- Universite de Versailles, 45 avenue des Etats-Unis, 78035 Versailles, France

[4]Microstructural Analysis Unit, UTS, Sydney, Australia

## Abstract

Optical properties of ZnMnO layers grown at low temperature by Atomic Layer Deposition and Metalorganic Vapor Phase Epitaxy are discussed and compared to results obtained for ZnMnS samples. Present results suggest a double valence of Mn ions in ZnO lattice. Strong absorption, with onset at about 2.1 eV, is tentatively related to Mn 2+ to 3+ photoionization. Mechanism of emission deactivation in ZnMnO is discussed and is explained by the processes following the assumed Mn 2+ to 3+ recharging.

# 1. Introduction

Alloys of ZnO and GaN with transition metal (TM) ions are promising materials for spintronics applications [1-3], due to the predicted their room temperature (RT) ferromagnetism (FM). It is now believed that RT FM, reported in several cases (see [2,3] and references given in), is due to inclusions of foreign phases of various TM oxides [2,4,5] or TM accumulation caused by a spinodal decomposition [5-9]. The latter predicted to occur inherently in diluted magnetic semiconductor (DMS) [6,7]. Fortunately, formation of foreign phases of $Mn_xO_y$ and of Mn accumulations can be depressed by lowering of the growth temperature of ZnMnO [4,10].

In this paper we compare optical properties of ZnMnO films obtained with two low temperature (LT) growth techniques: Atomic Layer Deposition (ALD) and Metalorganic Vapor Phase Epitaxy (MOVPE). The origin of an absorption, observed at "below band gap" spectral region in ZnMnO films is discussed. Based on the present investigations we propose the mechanism of photoluminescence (PL) deactivation in ZnMnO films.

# 2. Experimental
## 2.1. Samples

The present optical investigations were performed on ZnMnO epilayers grown at LT by ALD and MOVPE. In ALD we employed metalorganic zinc and manganese precursors: zinc acetate [$Zn(CH_3COO)_2$], and two types of Mn precursors: manganese-tris- 2,2,6,6-tetra methyl -3,5-heptanedione [denoted as $Mn(thd)_3$] and tris(2,4-pentanedionato) manganese(III) [denoted as $Mn(acac)_3$]. As an oxygen precursor we used deionized water vapor or ozone. Temperature of precursors were as follows: 230 – 250 ºC for zinc acetate, RT for water, 160 ºC for $Mn(acac)_3$, and 160-180 $^0$C for $Mn(thd)_3$. Substrate [(0001) sapphire or (0001) sapphire/GaN)] temperature was selected between 280 – 360 ºC. Pressure in a growth chamber, i.e., a pressure of $N_2$ transport and purging gas, was a few mbar. Ratio of Zn – to - Mn ALD cycles was either, 9 to 1 or 10 to 1, which we found to be optimal for a growth of fairly depth homogeneous ZnMnO films with low Mn fractions [10].

MOVPE samples were grown on (0001) sapphire or silica using DEZn and tertiary-butanol as zinc and oxygen precursors. Liquid tri-carbonyl-methylcyclopentadienyl-manganese [$(CO)_3CH_3C_5H_4Mn$] was used as manganese precursor and hydrogen as a vector gas. The substrate temperature was 450 $^oC$.

## 2.2 Optical set ups

The cathodoluminescence (CL) spectra were taken at room temperature (RT) with a JEOL 35C scanning electron microscope (SEM). A MonoCL2 CL system by Oxford Instruments was used for the CL detection. CL was collected by a parabolic mirror and injected to a monochromator equipped with a Hamamatsu R943-02 Peltier cooled photomultiplier. Spectra were not corrected for the system's response. Optical absorption investigations were performed at RT using Perkin Elmer Lambda 950 spectrophotometer. Since reflection of the studied films was low, the spectra were directly recorded in absorption mode. Further details on PL and CL set ups can be found in the reference [11].

## 3. Experimental results
## 3.1 Profiling of ZnO and ZnMnO emission with cathodoluminescence

CL investigations were performed at room temperature on a series of LT ZnO and ZnMnO layers grown by ALD on either sapphire or sapphire/GaN substrates with Mn content up to 18 %. In the case of ZnMnO layers we failed in most of the cases to observe a visible light emission. The CL emission was observed only for the samples with the low Mn fractions as described recently [11].

Detail depth-profiling investigations were done for the film grown with nearly equal sequences of zinc and manganese cycles (2 to 1), which, from the SIMS investigations discussed elsewhere [4,10], we found to be fairly depth inhomogeneous. For these samples we observed Mn accumulation at surface of films, i.e. deeper regions of samples were relatively Mn poor [4,10]. Once Mn concentration was homogenized, by selection of the proper zinc-to-Mn ALD sequences and the pulsing and purging times [10], the CL emission was not observed (for higher Mn content) or was very weak.

The CL data shown in Fig. 1 (a, b) were taken at different voltages, using the so-called depth-profiling option of the CL method. In the CL depth-profiling, by changing accelerating voltage, we can evaluate depth homogeneity of light emission. Details of the method can be found elsewhere, for example in our recent papers on GaN [12,13].

The CL data shown in Fig. 1 (a) were taken for the as-grown ZnO film, which was deposited at LT (using zinc acetate) on sapphire substrate covered with 1 μm thick GaN layer grown by MOVPE. The „band edge" emission, of excitonic character, dominates the RT CL spectrum. This emission is observed together with a defect-related red CL, discussed in several reports, as summarized recently in [14]. Intensities of the both emissions increase with excitation moved from an interface region, i.e. at reduced accelerating voltages. The defect-related red emission is relatively strongest at the interface region, indicating strong defecting of this region. Such depth property of the CL emission we commonly observed for GaN-based structures [12,13].

The observed depth properties of ZnMnO emission are shown in Fig. 1 (b). The quite different situation is observed. Now, the CL emission mostly comes from deeper regions of the sample and is relatively weak, when excited from the surface-close area of the sample. The "band edge" CL is relatively (as compared to the defect-related red CL) the weakest at the surface close area of the film, i.e. from the Mn rich region. The width of the „band edge" CL is not affected by a variation of excitation conditions, which indicates that the observed changes in the CL intensity are related to depth variations in efficiency of nonradiative recombination. No emission due to $Mn^{2+}$ ions was detected.

### 3.3 Optical absorption investigations

For ZnMnO samples we consistently observed an absorption band appearing below the onset of the band-to-band transition. This band, shown in Figs. 2 and 3, depends on Mn fraction in the sample, as we concluded from the optical absorption measurements performed for the MOVPE-grown ZnMnO layers with different Mn fractions. The additional absorption is so strong in samples with larger Mn fractions that it masks onset of the fundamental band gap absorption.

This Mn-related absorption band was first reported by Fukumura et al. [15] and was tentatively related to charge transfer transitions between donor and acceptor ionization levels of Mn ions and the band continuum [15]. In other reports this band was related to smeared out the $Mn^{2+}$ intra-shell transitions [16].

In Fig. 3 we compare the $Mn^{2+}$ intra-shell PL and PL excitation (PLE) observed by us in ZnMnS bulk sample and for ZnMnS nanopowders [17] (Fig. 3 (a)) with the absorption seen by us for two types [grown with $Mn(thd)_3$ and $Mn(acac)_3$] of our LT (at 280 $^o$C) ZnMnO films, grown by the ALD (Fig. 3 (b)). For this study we used lightly Mn doped films [about 1 % for the film grown with $Mn(acac)_3$, sample #268, and about 3 % for film grown with $Mn(thd)_3$, sample #273].

In ZnMnS several intra-shell transitions are resolved for samples with different Mn fractions [17]. It is thus unlikely that intra-shell transitions are smeared out in ZnMnO layers. We must thus look for other explanations of their absence.

Observation of these transitions in ZnMnS allows us to verify critically the accuracy of the cluster model calculations presented in the reference [16]. Energy position of $^6A_1$ to $^4T_1$, $^4T_2$, $^4E$ and $^4A_1$ transitions were calculated, using the cluster model to be (in eV): 2.31, 2.57, 2.70 and 2.76 (indicated with the arrows in Fig. 3 (a, b)) for ZnMnS and 2.55, 2.85, 2.97 and 2.99 for ZnMnO.

The accuracy of calculations is limited (see Fig. 3 (a)). The transition energies are overestimated by about 0.05 - 0.1 eV in the case of ZnMnS. We can, however, estimate that the $^6A_1$ to $^4T_1$ transition in ZnMnO should occur at about 2.50 - 2.45 eV and that to $^4T_2$ level at 2.80 - 2.75 eV. No extra absorption is seen at these energies in the absorption spectrum shown in Fig. 3 (b). Instead a broad absorption band is seen, which clearly depends on Mn content in the sample. We will speculate further on that this broad absorption is due to a Mn-related photo-ionization transition.

## 4. Discussion

PL of ZnMnO was either not observed or consisted of a broad emission [18] or of a weak excitonic lines [19,20]. In all the cases the PL was rather weak, which led to the opinion that Mn acts as a "killer" of a visible PL in ZnO [11,20]. The $Mn^{2+}$ intra-shell PL was not observed.

A possible explanation of ZnMnO PL properties comes from the investigations of GaMnN (see [2] and references given there). Multivalence of Mn ions in GaN lattice was quite convincingly proved [21-25]. $Mn^{2+}$ to $Mn^{3+}$ recharging was observed upon illumination in GaMnN [21]. It was also claimed that multivalence of Mn in GaMnN can influence magnetic properties of this material [22]. $Mn^{4+}$ to $Mn^{3+}$ recharging was also evidenced there [23,24], but its origin remains unclear. It may as well be due to existence of foreign phases in GaMnN [25].

If we take 1.8 eV as 2+ to 3+ ionization energy in GaN [30], 1.3 eV as the value of ZnO/GaN conduction band alignment [27], and use the concept of an universal reference level [20], we can roughly estimate the relevant ionization energy in the ZnO lattice to be about 0.4 eV below the edge of the conduction band of ZnO. However, the model of the universal reference level works well only in the case of systems of the same anion, and predictions of this model are less accurate once only cation is the same [28]. Here we deal with the two materials (GaN and ZnO) of a different anion and cation, and a different ionicity, so the accuracy of the method is limited.

If we attribute the observed absorption, with the onset at about 2.0 - 2.1 eV, to the 2+ to 3+ Mn ionization in ZnO, the results of optical investigations can be consistently interpreted. Assuming the photoionization character of the absorption shown in Figs. 2 and 3, we can evaluate, using Kopylov and Pihktin approach [29], the relevant ionization energies. In this model the optical cross section σ(E) is given by:

$$\sigma = const \int_{-\beta}^{\infty} dz\, e^{-z^2} \sigma_{el}(E_{opt}, E + \Gamma z)(1 + \frac{\Gamma z}{E})$$

where $\sigma_{el} \propto \dfrac{(E - E_{opt})^{3/2}}{E^3}$ is a purely electronic optical cross section and β and γ are given by:

$$\beta = \frac{(E - E_{opt})}{\Gamma}$$

$$\Gamma = \frac{\hbar\omega_{exc}}{\hbar\omega_0} \sqrt{2(E_{opt} - E_{th})\hbar\omega_0 \coth \frac{\hbar\omega_0}{2kT}}$$

and Γ describes a spectrum broadening caused by coupling to the lattice vibrations. $\hbar\omega_{exc}$ and $\hbar\omega_o$ are vibronic energies in the ground and excited states, respectively, which commonly are set to be equal, T is the temperature, and k is the Boltzmann's constant.

In the model photoionization energies of impurities are described using two parameters – optical ionization energy ($E_{opt}$) and thermal ionization energy ($E_{th}$), with their difference equal to a lattice relaxation energy ($E_{rel}$). Using the above formulas we get $E_{opt}$ = 2.6 ± 0.1 eV and $E_{th}$ = 2.1 ± 0.1 eV.

If the band shown in Figs. 2 and 3 is due to Mn 2+ to 3+ ionization, we can explain many of the puzzling optical properties of ZnMnO, in particular why the $Mn^{2+}$ intra-shell transitions are missing both in absorption and in PL experiments. The value of the lattice relaxation energy (about 0.5 eV) is too small to introduce a significant barrier for a fast autoionization from impurity excited states overlapping with the conduction band states (see Fig. 4 (a)). Similar situation we found previously for the intra-shell transitions of $Fe^{2+}$ ion in ZnS [30]. Only in the case of large lattice relaxation energies, expected e.g. in more ionic crystals, intra-shell transitions can still be resolved, even though they overlap with photoionization transitions. Such situations (Fig. 4 (b)) we observed previously for the $Eu^{2+}$ intra-shell transitions in $CdF_2$ crystals [31].

Finally, we should account for the observed PL deactivation by Mn ions in ZnO lattice. The explanation is based on the results of our previous investigations of PL deactivation in ZnS:Fe [30,32]. There PL quenching was related to the three mechanisms: 1) efficient carrier recombination via a mid band gap Fe-related level (the so-called bypassing effect [32]); 2) efficient Auger-type energy transfer from excitons and DAPs to Fe ions, and; 3) formation of complex centers of Fe with common PL activators in wide band gap II-VI materials. First two of these processes (bypassing and the Auger effect) are most likely responsible for the PL deactivation in the present case, assuming Mn ionization.

**Conclusions**

Concluding, if we assume the mix valence of Mn ions in ZnMnO, we can account for the lack of Mn intra-shell transitions and for the PL quenching. If so, it looks very unlikely that we can simultaneously realize a required high concentration of Mn ions in 2+ charge state and p-type conductivity to achieve RT FM.

**Acknowledgments**

This work was partly supported by grant no. 1 P03B 090 30 of MEiSW granted for the years 2006-2008 and by FunDMS ERC Advanced Grant Research.

**Figure captions**:

**Figure 1 (a, b)**: Depth-profiling CL properties of (a) ZnO and (b) ZnMnO observed at RT for LT ZnO, ZnMnO layers grown by ALD.

**Figure 2**: Comparison of absorption spectra in bulk (about 5 % of Mn) and in LT MOVPE layers of ZnMnO with Mn fractions 1, 4 and 22 %. The spectra were normalized in the way that 1 in the normalized absorption stands for 0% of the transmission.

**Figure 3**: Comparison of PL and PLE spectra for ZnMnS (a) with absorption spectrum for two types of ALD ZnMnO samples [sample #268 grown with Mn(acac)$_3$ and sample #273 grown with Mn(thd)$_3$] (b).

**Figure 4**: Configuration coordinate model of Mn$^{2+}$ ionization and intra-shell transitions in ZnMnO (a) and in a hypothetical ionic system with a large lattice relaxation energy (b). In the latter case intra-shell transitions should be resolved despite their energy degeneration with continuum of the conduction band.

**Fig. 1**

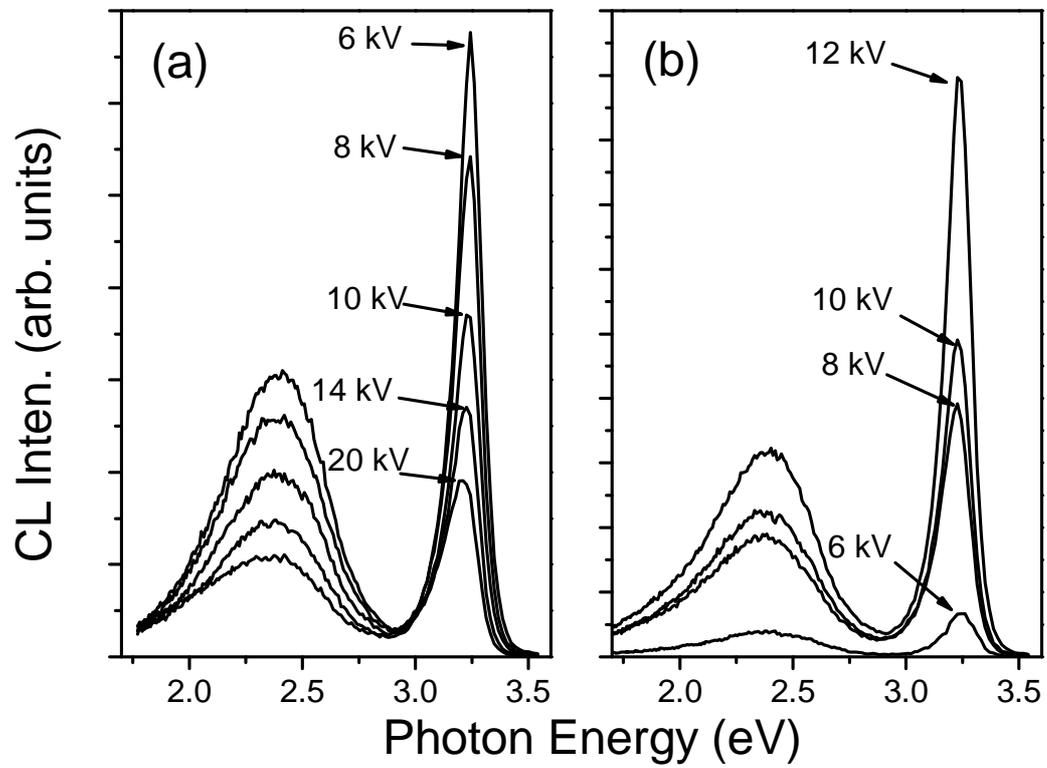

**Figure 1 (a, b)**: Depth-profiling CL properties of (a) ZnO and (b) ZnMnO observed at RT for LT ZnO, ZnMnO layers grown by ALD.

**Fig.2**

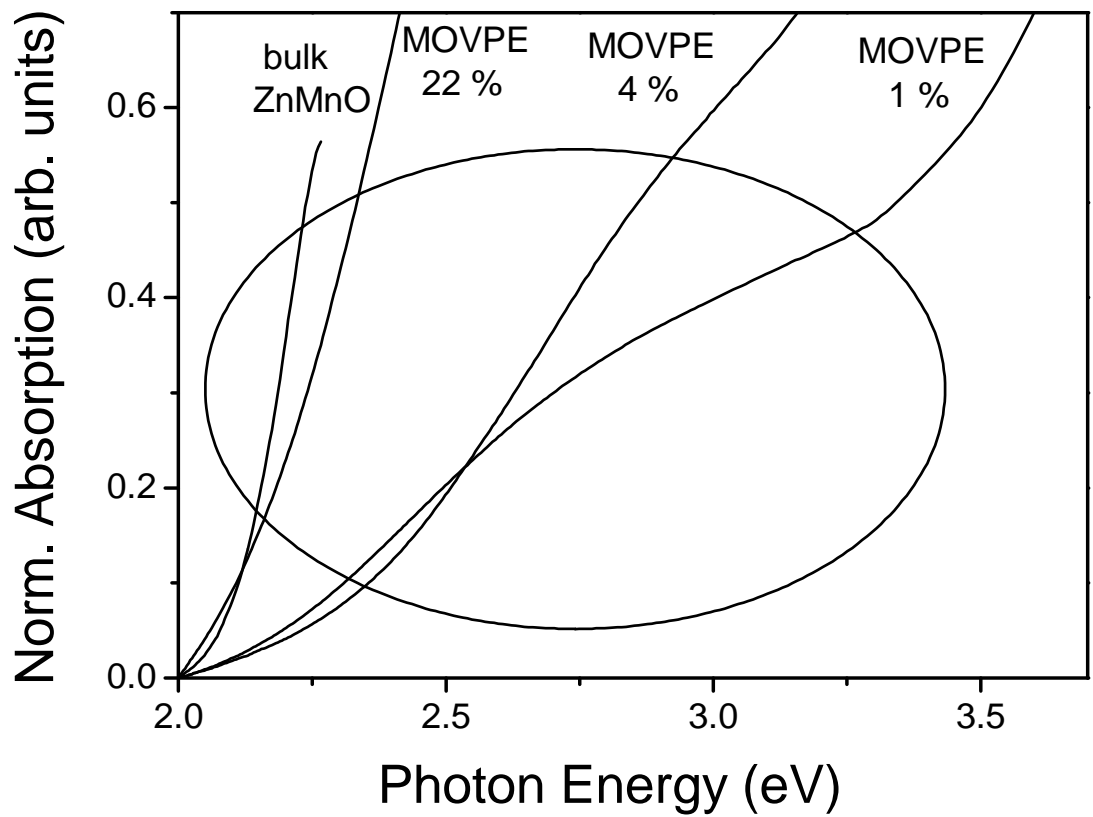

**Figure 2**: Comparison of absorption spectra in bulk (about 5 % of Mn) and in LT MOVPE layers of ZnMnO with Mn fractions 1, 4 and 22 %. The spectra were normalized in the way that 1 in the normalized absorption stands for 0% of the transmission.

**Fig.3**

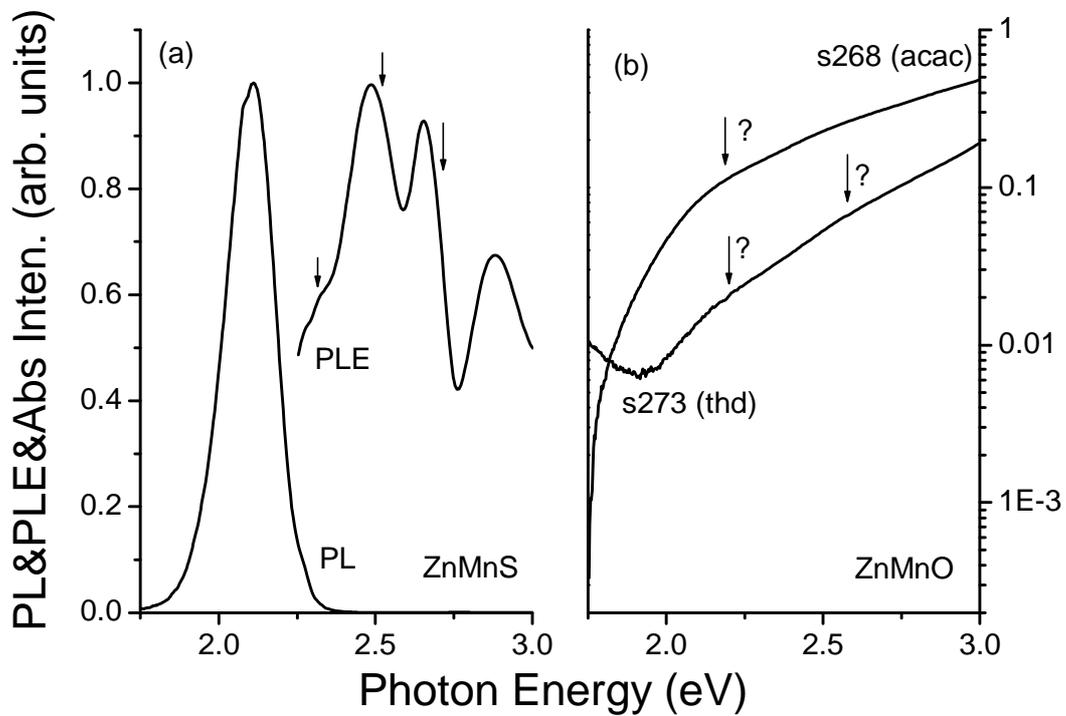

**Figure 3**: Comparison of PL and PLE spectra for ZnMnS (a) with absorption spectrum for two types of ALD ZnMnO samples [sample #268 grown with Mn(acac)$_3$ and sample #273 grown with Mn(thd)$_3$] (b).

**Fig.4**

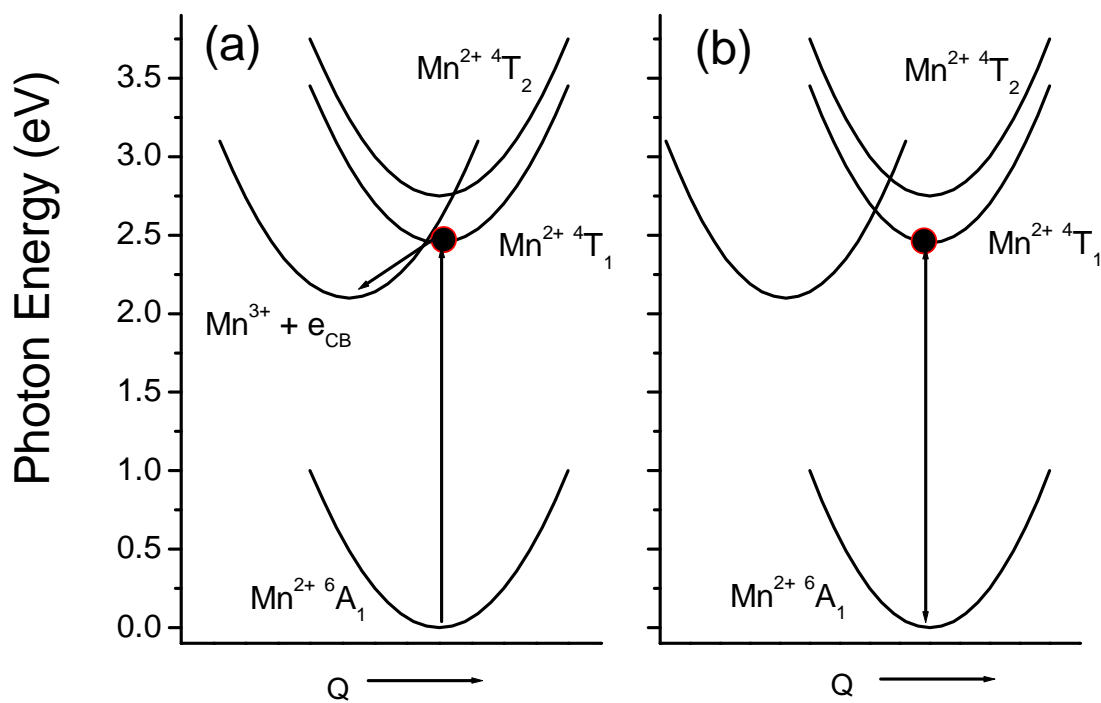

**Figure 4**: Configuration coordinate model of $Mn^{2+}$ ionization and intra-shell transitions in ZnMnO (a) and in a hypothetical ionic system with a large lattice relaxation energy (b). In the latter case intra-shell transitions should be resolved despite their energy degeneration with continuum of the conduction band.